# On-axis diffraction-limited design of bi-parabolic singlet lenses

Juan Camilo Valencia-Estrada[1] and Jorge García-Márquez[2*]

[1] Oledcomm, 10-12, avenue de l'Europe, Vélizy-Villacoublay, 78140, France

[1]E-mail: juanvalenciaestrada@gmail.com

*[2]E-mail: jorge.garcia.marquez@gmail.com

**Abstract**

A lens limited by diffraction and having two parabolic surfaces is presented. The knowledge of the following parameters: object distance, relative refractive index, lens thickness, and image distance, enables to analytically calculate the parabolic front and back surfaces. The conditions to obtain diffraction-limited images for these lenses and its maximal diameter are described. These bi-parabolic lenses can be easily manufactured at reduced costs and can be used for several commercial and industrial applications. The method to obtain such surfaces and a simple example validated by using Oslo® are described here.

## 1. Introduction

A paraboloid of revolution concentrates in its focus the light from an object placed at infinity, this is a reason for astronomers to use them as primary mirrors [1,2]. Nevertheless, parabolic refractive interfaces are not common; this is probably due to the fact that they introduce an important amount of spherical aberration. Concerning ophthalmic use of paraboloids, it can be said that the eye lens is frequently modelled as having parabolic interfaces [3,4]. Not only the eye-lens but also the base curve for some contact lenses has parabolic interface surfaces [5]. In some cases, it has been recommended to use parabolic interfaces to reduce spherical aberration [6]. Flat-parabolic lenses are also used in X-rays equipment [7] and in integrated optics [8-9].

Recently, Lozano et al. [10] have found a method to correct spherical aberration introduced by a parabolic surface by using a correcting surface. Given any real object and real image, Gonzalez et al. have shown a different method to find an aspherical corrective posterior surface with simplified signs' rules [11].

Parabolic lenses have a simpler mathematical representation than spherical or non-parabolic aspherical lenses. Any parabolic lens is a special case of an aspheric lens, but its simpler mathematical representation makes a parabolic lens easier to use. Nevertheless, it is not common to find lenses having both surfaces parabolic. Recently a bi-parabolic catadioptric lens used as optical link element for visible light communication has been proposed [12]. Even more, it is less common to find bi-parabolic lenses limited by diffraction. As a matter of fact, the authors have not found any work published or patented concerning them.

Here, a method to easily calculate the best approximation [13] of any correcting surface to a paraboloid is found. A described back corrective aspheric surface with any conical constant and deformation coefficients can be fitted to a paraboloid surface by making its conical constant [14] $K = -1$ and vanishing all the deformation coefficients. Thus, by equalling a fourth degree coefficient to zero, and due the fact the series converges very fast, it is possible to find an approximation to the best combination of parabolic surfaces. It has been assumed that coefficients equal or higher than 6th degree do not have a significant impact in the diffraction limited image. Thus, a bi-parabolic lens may reduce dramatically spherical aberration below the diffraction limit.

## 2. Modelling rays

The coordinate's origin is set at the front or anterior surface vertex, and sub index $a$ (for anterior) will be used. The anterior surface $z_a$ is a revolution paraboloid whose meridional section is represented in cylindrical coordinates by

$$z_a(r) = \frac{r^2}{4f_a} = \frac{r^2}{2R_a} = \frac{c_a r^2}{2} = \frac{c_a r^2}{1+\sqrt{1-(1+K_a)c_a^2 r^2}} \; , \quad (1)$$

where $r$ is the abscissa, $f_a$ is the anterior geometrical –reflective– focal distance [10]. From this equation, the focal distance $f_a$ is assumed as a geometric characteristic of any parabola that, according with the definition, is a set of points equidistant between a focus and a bisector; $R_a$ is the curvature radius in its vertex, with $R_a = 2f_a$, and $c_a$ is the vertex curvature and $K_a = -1$ its conic constant.

The posterior surface, represented by sub-index $b$ can correct all spherical aberration orders and can be represented by

$$\hat{z}_b(r) = t + \frac{c_b r^2}{1+\sqrt{1-(1+K_b)c_b^2 r^2}} + \sum_{j=2}^{\infty} B_{2j} r^{2j} \; , \quad (2)$$

having a vertex curvature $c_b$, a conical constant $K_b$ and an infinite number of deformation coefficients $B_{2j}$. According to Eq. (15) in [10], the back correcting surface has a vertex curvature radius expressed by

$$c_b = \frac{(n-1)(t+nt_b)t_a + R_a[t+n(t_b-t_a)]}{(n-1)t_b[R_a(nt_a-t)-(n-1)tt_a]} = \frac{(n-1)(t+nt_b)t_a + 2f_a[t+n(t_b-t_a)]}{(n-1)t_b[2f_a(nt_a-t)-(n-1)tt_a]}. \quad (3)$$

Here, $n$ is the relative lens refraction index, $t_a$ is the object distance, $t$ is the central lens thickness, and $t_b$ the image distance as depicted in Fig. 1. Only lenses immersed in the same refractive medium are considered here. A different case is out of scope in this work.

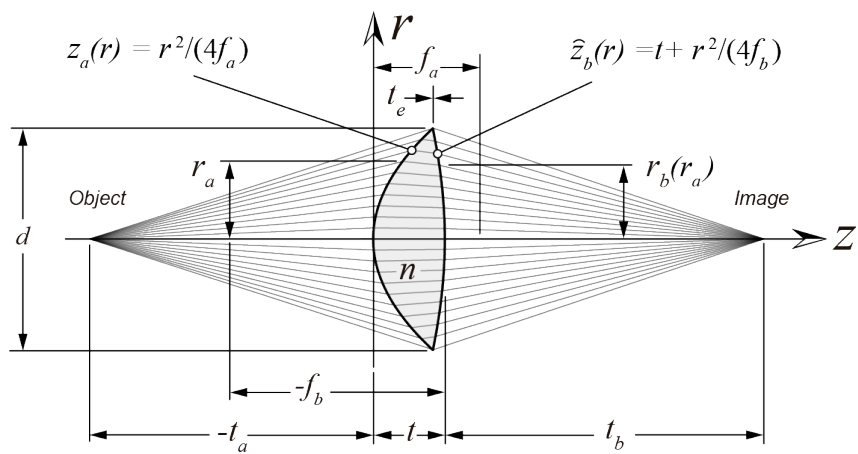

**Figure 1.** Bi-parabolic diffraction-limited lens.

According to Eqs. (5.7 to 5.12b) in reference [13], the deformation coefficients $B_{2j}$ are calculated analytically as functions of the conic constant $K_b$.

$$B_4(K_b) = A_4 - \left(\frac{c_b}{2}\right)^3 (1+K_b),$$

$$B_6(K_b) = A_6 - 2\left(\frac{c_b}{2}\right)^5 (1+K_b)^2,$$

$$B_8(K_b) = A_8 - 5\left(\frac{c_b}{2}\right)^7 (1+K_b)^3, \quad (4)$$

$$B_{10}(K_b) = A_{10} - 14\left(\frac{c_b}{2}\right)^9 (1+K_b)^4,$$

$$\vdots$$

$$B_{2j}(K_b) = A_{2j} - \frac{2^{2j-1}\text{Binomial}(1/2, j)}{(-1)^{j+1}} \left(\frac{c_b}{2}\right)^{2j-1} (1+K_b)^{j-1},$$

whose coefficients $A_{2j}$ are calculated by implicitly derivating the parametric correcting surface $[r_b(r_a), z_b(r_a)]$ [11]:

$$\left.\begin{aligned} r_b(r_a) &= r_a + (\alpha\chi)|_{r_a}, \\ z_b(r_a) &= \frac{r_a^2}{4f_a} + (\beta\chi)|_{r_a}, \end{aligned}\right\} \quad (5)$$

with director cosines $\alpha$ and $\beta$ of the ray travelling inside the lens

$$[\alpha, \beta] = \left[\frac{2f_a r_a(8f_a^2 - 4f_a t_a + r_a^2)}{\delta} - r_a\phi, \frac{r_a^2(8f_a^2 - 4f_a t_a + r_a^2)}{\delta} + 2f_a\phi\right], \quad (6)$$

and the following recurrent variables: $\chi$ is a special function that depends on the parametric abscissa $r_a$,

$$\chi = \frac{\varphi}{\varepsilon} - \frac{1}{\varepsilon}\sqrt{\varphi^2 + \varepsilon\left\{r_a^2 + (t+t_b)^2 - \frac{r_a^2[8f_a(t+t_b) - r_a^2]}{16f_a^2} - \frac{(\varphi-\gamma)^2}{n^2}\right\}}, \quad (7)$$

with

$$\left.\begin{aligned} \delta &= n(4f_a^2 + r_a^2)\sqrt{r_a^2 + 8f_a\left[2f_a\left(r_a^2 + t_a^2\right) - r_a^2 t_a\right]}, \\ \varepsilon &= n^2 - 1, \\ \phi &= \frac{1}{\delta}\sqrt{\varepsilon r_a^6 - 32(\varepsilon-1)f_a^3 r_a^2 t_a - 8\varepsilon f_a r_a^4 t_a + 64 f_a^4(\varepsilon r_a^2 + n^2 t_a^2) + 4f_a^2\left[(1+5\varepsilon)r_a^4 + 4\varepsilon r_a^2 t_a^2\right]}, \\ \gamma &= \alpha r_a - \beta(t+t_b - r_a^2/4f_a), \quad \text{and} \\ \varphi &= \gamma + n\left[nt - t_a + t_b - \sqrt{r_a^2 + (r_a^2/4f_a - t_a)^2}\right]. \end{aligned}\right\} \quad (8)$$

Equations (5-8) are valid for any real object and real image.

Coefficients $A_{2j}$ were calculated following the procedure in reference [13] and its Eqs. (5.7- 5.12b):

$$A_2 = \frac{c_b}{2} = \frac{1}{2}\lim_{r_a \to 0}\left[\frac{d^2 z_b/dr_a^2}{(dr_b/dr_a)^2}\right],$$

$$A_4 = -\frac{c_b}{6}R_3 + \frac{1}{24}Z_4 = \lim_{r_a \to 0}\left[-\frac{1}{6}\left(\frac{d^2 z_b/dr_a^2}{(dr_b/dr_a)^2}\right)\left(\frac{d^3 r_b/dr_a^3}{(dr_b/dr_a)^3}\right) + \frac{1}{24}\left(\frac{d^4 z_b/dr_a^4}{(dr_b/dr_a)^4}\right)\right], \quad (9)$$

$$\vdots,$$

the deformation coefficients $A_{2j}$ in Eq. (9) are obtained in the limit when the abscissa $r_a$ tends to zero; the coefficient $A_2$ has been used to obtain $c_b$ in Eq. (3). Here, the coefficients are represented in their calculus limit form to save space.

Derivatives are evaluated in the surface's vertex. $B_{2j}$ coefficients are reduced to $A_{2j}$ coefficients using a paraboloid as the posterior surface to establish a back parabolic surface base. In this case the conic constant $K_b = -1$ has been chosen, so Eq. (2) is reduced to

$$\hat{z}_b(r) = t + \frac{r^2}{4f_b} + \sum_{j=2}^{\infty} A_{2j} r^{2j}, \quad (10)$$

and the back geometrical focal distance is obtained by simplifying Eq. (3)

$$f_b = \frac{(n-1)t_b[2f_a(nt_a - t) - (n-1)tt_a]}{2\{(n-1)(t + nt_b)t_a + 2f_a[t + n(t_b - t_a)]\}}. \quad (11)$$

Any back corrective aspheric surface having any conical constant and deformation coefficients can be fitted to a paraboloid surface by making its conical constant $K_b = -1$ and vanishing all the deformation coefficients. Thus, by equalling $A_4$ to zero, and due to the series' very fast convergence, it is possible to find an approximation to the optimal surface combination, by equalling coefficient $A_4$ to zero

$$A_4 = -\frac{c_b}{6}R_3 + \frac{1}{24}Z_4 = 0. \quad (12)$$

Following Eq. (9), is obtained

$$\lim_{r_a \to 0}\left[4\left(\frac{d^2 z_b}{dr_a^2}\right)\left(\frac{d^3 r_b}{dr_a^3}\right) - \left(\frac{d^4 z_b}{dr_a^4}\right)\left(\frac{dr_b}{dr_a}\right)\right] = 0. \quad (13)$$

Thus, the next fourth degree polynomial for the unknown variable $f_a$ is obtained

$$pU^4 + 2nVU^3 t_b - 2nU^2 V^2 t_b^2 + nt_b^3\left\{16n^2 f_a^3 t_a\left[(n^2 + n - 2)t_a - (n^2 - n - 2)f_a\right] + 8n^2 m^2 f_a^2 t_a^3 - ptV^4\right\} = 0, \quad (14)$$

having the following recurrent variables

$$\left.\begin{array}{l} m = n - 1, \\ p = n + 1, \\ V = t - nt_a, \\ U = mtt_a - 2f_a V, \\ W = 2f_a + mt_a. \end{array}\right\} \quad (15)$$

The solution for $f_a$ can be obtained analytically or numerically. This quartic Eq. (14) has two real and two complex roots. Both real solutions may be optically valid for any transparent material: one for lenses with positive magnification, and the other for lenses with negative magnification. Depending on the combination of conjugated plane distances, the real solution can be obtained following the next steps:

1. Using the set of recursive variables

$$\left.\begin{aligned}
\hat{a} &= pV^4 + 2nt_bV^3 - 2nt_b^2V^2 + npt_b^3(n(m-1)V - m^2t), \\
\hat{b} &= \frac{m^4tt_a^4}{16\hat{a}}\Big(p(t^3 - nt_b^3) + 2ntt_b(t - t_b)\Big), \\
\hat{c} &= \frac{m^3tt_a^3}{4\hat{a}}\Big(2pt^2V^2 + nt(t + 3V)t_b - 2n(t+V)t_b^2 - 2npt_b^3\Big), \\
\hat{d} &= \frac{m^2t_a^2}{2\hat{a}}\Big(3pt^2V^2 + 3nt(t+V)Vt_b - n\big((2t+V)^2 - 3t^2\big)t_b^2 + n(n^2t_a - 3pt)t_b^3\Big), \\
\hat{e} &= \frac{mt_a}{\hat{a}}\Big(2ptV^3 + nV^2(3t+V)t_b - 2n(t+V)Vt_b^2 + n\big((n+2)n^2t_a - 2pt\big)t_b^3\Big), \\
\hat{f} &= \hat{b} - \frac{\hat{c}\hat{e}}{4} + \frac{\hat{d}\hat{e}^2}{16} - \frac{3\hat{e}^4}{256}, \\
\hat{g} &= \hat{c} - \frac{\hat{d}\hat{e}}{2} + \frac{\hat{e}^3}{8}, \\
\hat{h} &= \hat{d} - \frac{3\hat{e}^2}{8}.
\end{aligned}\right\} \quad (16)$$

2. By doing so, and taking the variable change $\{f_a \to W - \hat{e}/4\}$, the equation (14) can be expressed as a new fourth degree equation with no third degree term

$$\hat{f} + \hat{g}W + \hat{h}W^2 + W^4 = 0, \tag{17}$$

Again, using Descartes method, equation (17) can be factored from a fourth degree equation into two second degree equations having real roots

$$\hat{i} + \hat{j}W + W^2 = 0, \tag{18}$$

whose coefficients $\hat{i}$ and $\hat{j}$ are obtained with the real solution of

$$\left.\begin{aligned}
\hat{h} &= \hat{i} + \frac{\hat{f}}{\hat{i}} - \hat{j}^2 = 0, \\
\hat{g} &= \hat{j}\left(\frac{\hat{f}}{\hat{i}} - \hat{i}\right) = 0
\end{aligned}\right\}, \tag{19}$$

3. Finally, the anterior geometrical focal distance can be calculated by using

$$f_a = W - \frac{\hat{e}}{4}, \tag{20}$$

4. Once the anterior focal distance is calculated the geometrical back focal distance is found by using Eq. (11),

$$f_b = \frac{-mUt_b}{2(U + nWt_b)}. \tag{21}$$

This procedure is easily assessed by ray tracing.

## 3. Maximum aperture diameter

The lens focal distance $F$ is a paraxial characteristic of a lens. The lens maker's formula is a simple way to calculate it

$$\frac{1}{F} = (n-1)\left(\frac{1}{R_a} - \frac{1}{R_b} + \frac{(n-1)t}{n R_a R_b}\right) = (n-1)\left(\frac{1}{2f_a} - \frac{1}{2f_b} + \frac{(n-1)t}{4n f_a f_b}\right), \quad (22)$$

here, $R_b = 1/c_b$ is the curvature radius at the back parabolic interface vertex. An alternative formula is Gullstrand's

$$\frac{1}{F} \cong -\frac{1}{t_a} + \frac{1}{t_b} + \frac{t}{n\, t_a\, t_b}. \quad (23)$$

To calculate the lens aperture that allows us to obtain a diffraction limited image, a marginal ray in the meridional plane is traced to calculate the height $h(d/2)$ on the image plane. Considering that marginal rays are the most aberrated rays, thus, the following condition can be proposed

$$|h(d/2)| \leq 1.22\lambda |F/\#| \quad \text{or} \quad d|h(d/2)| \leq 1.22\lambda |F|, \quad (24)$$

to numerically find the maximum aperture diameter $d$.

After vanishing all the deformation coefficients in the sum of Eq. (10) the back surface is

$$\hat{z}_b(r) = t + \frac{r^2}{4 f_b} \quad (25)$$

Finally, the edge thickness $t_e$ is calculated with the condition

$$t_e = \hat{z}_b(d/2) - z_a(d/2) > 0. \quad (26)$$

For a bi-parabolic lens design, Eq. (26) can be reduced to

$$t_e = t + \frac{d^2}{16 f_b} - \frac{d^2}{16 f_a} > 0. \quad (27)$$

## 4. Lens design example

A bi-parabolic diffraction-limited lens to exemplify this method is explained here below. The lens has the following input parameters: (a) Object distance $t_a = -800$ mm, (b) lens central thickness $t = 0.6$ mm, (c) image distance $t_b = 12$ mm, (d) diameter $d = 5$ mm, its material is plastic with a refraction index $n = 1.76$ at $\lambda = 589$ nm. Focal distances are calculated by using equations (20-21); then, $f_a = 5.0973$ mm and $f_b = -46.594$ mm are obtained. These focal distances corresponds to the curvature radii $R_a = 10.1946$ mm and $R_b = -93.188$ mm respectively. By introducing these values in any lens design software it is easy to verify that for a $F/\# = 2.46$ the image is diffraction limited [Fig 2(b)].

The reader can easily assess this example with any optical design software just by following the next steps: i) to compute $f_a$ from equation (20) using the equations (14-20); ii) to compute $f_b$ from Eq. (21); iii) to calculate the focal distance $F$ from equation (22); iv) the maximum aperture diameter is determined by using Eq. (23), this ensures a diffraction limited image. v) The edge thickness $t_e$ should meet Eq. (27). Nevertheless, if $t_e < 0$, the central thickness must be increased which implies the repetition of all the steps.

A change of the optical material implies a change in the spot diagram. Now, consider that a modified Poly-Methyl-Methacrylate (PMMA) is used. Its refraction index is $n=1.4875$ at $\lambda=589$ nm; the focal distance calculation results in $f_a = 3.68539$ mm and $f_b = -14.596$ mm. These distances correspond to $R_a = 7.37079$ mm and $R_b = -29.192$ mm curvature radii respectively. In figure 2(c) it can be seen that the image is not diffraction limited; its spot diagram for a $F/\# = 2.46$ was calculated. In spite of its low refraction index this lens can be taken to diffraction limit by reducing its diameter to $d = 4.2$ mm. This results in an increase of its $F/\#$ to 2.93, as observed in Fig. 2(d).

## 5. Discussion and conclusions

The method here exposed entails four main advantages: i) it permits to obtain a diffraction limited image. This advantage is due to the fact the approximation in Eq. (5) has a deformation series that converges very fast, ii) it permits to simplify the design process of basic optical systems, iii) bi-parabolic lenses can be used to design systems where an image (diffraction-limited) point can be used as an object point by the following element. As a matter of fact every combination of conjugated plane distances has a unique optimal design; finally iv) it permits to mathematically represent a lens in its simplest manner. The simulation conducted here has also permitted to conclude that coma aberrations are generally lower than coma in spherical lenses as it can be easily assessed by using any commercial optical design software. Nevertheless, field curvature and astigmatism may increase following an applicative design. The method assessment has been performed by using Oslo® for the ray tracing incurring neither in optimization nor in defocus.

Aberrations theory for this kind of bi-parabolic lenses is an open domain.

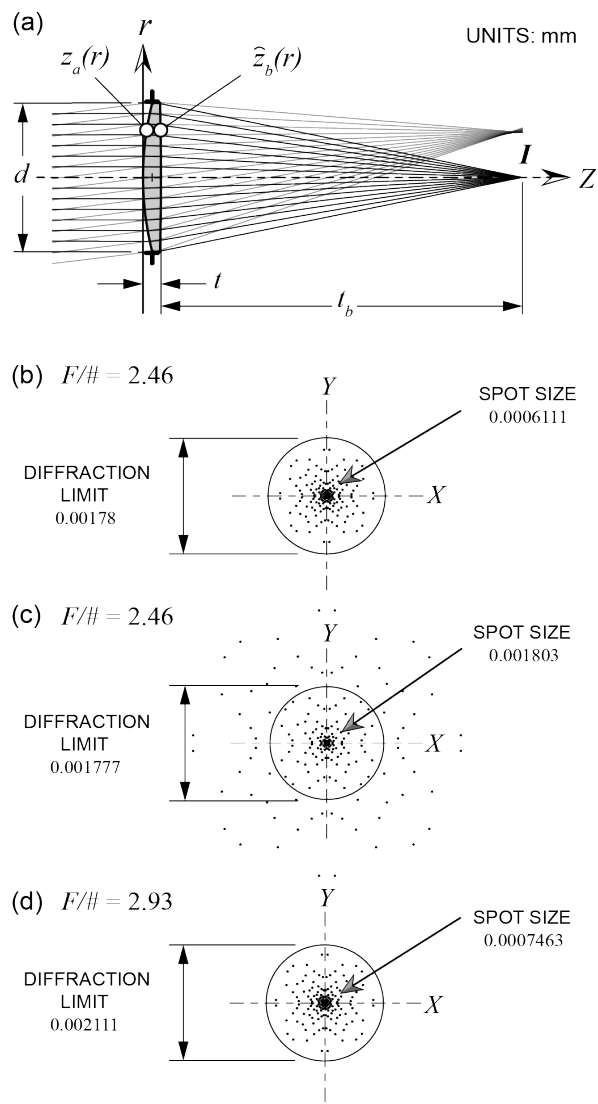

**Figure 2**. (a) Ray tracing for two different beams. The object height is −100 mm and the Gaussian image height is 1.538442 mm. (b) In this first example that corresponds to the depicted ray tracing $n = 1.76$; (c) by changing the refractive index to $n = 1.4875$ the image deteriorates, nevertheless it can be corrected by reducing the lens diameter as shown in (d). Geometrical spot size and diffraction limit are in mm. $\lambda = 589$ nm. These results were assessed in Oslo®.

## Acknowledgements

The results here obtained were possible thanks to the funding received from the European Union's Horizon 2020 Research and Innovation Actions programme under grant agreement No. 761992.

## References

[1] C. W. Akerlof, R. L. Kehoe, T. A. McKay, E. S. Rykoff, D. A. Smith, D. E. Casperson, K. E. McGowan, W. T. Vestrand, P. R. Wozniak, J. A. Wren, M. C. B. Ashley, M. A. Phillips, S. L. Marshall, H. W. Epps, J. A. Schier, "The ROTSE-III Robotic Telescope System", Publications of the *Astronomical Society of the Pacific* **115**(803) (2003)132-140.

[2] H. M. Martin, J. H. Burge, D. A. Ketelsen, S. C. West, "Fabrication of the 6.5 m primary mirror for the multiple mirror telescope conversion", Proc. SPIE **2871** (1996) 399-404. https://doi.org/10.1117/12.269063

[3] R. Navarro, J. Santamaría, J. Bescós, "Accommodation-dependent model of the human eye with aspherics", J. Opt. Soc. Am. A **2**(8) (1985) 1273-1281. https://doi.org/10.1364/JOSAA.2.001273

[4] A. C. Kooijman, Light distribution on the retina of a wide-angle theoretical eye, J. *Opt. Soc. Am.* **73**(11) (1983) 1544-1550. https://doi.org/10.1364/JOSA.73.001544

[5] O. Wichterle, Method of production of plastic lenses with aspherical surfaces, Ceskoslovenska akademie, Prague (1970) Patent 3497577.

[6] M. Laikin, *Lens Design*, 4th Ed., CRC Press, Boca Raton, 2006.

[7] B. Lengeler, C. Schroer, J. Tümmler, B. Benner, M. Richwin, A. Snigirev, I. Snigirevab, M. Drakopoulos, "Imaging by parabolic refractive lenses in the hard X-ray range", *J. Synchrotron Rad.* **6** (1999) 1153-1167. https://doi.org/10.1107/S0909049599009747

[8] A. M. Rashed, K.A. Williams, P. J. Heard, R. V. Penty, I. H. White, "Tapered waveguide with parabolic lens: theory and experiment", *Opt. Eng.* **42**(3) (2002) 792-797. https://doi.org/10.1117/1.1541621

[9] M. Testorf, J. Jahns, Imaging properties of planar-integrated micro-optics, *J. Opt. Soc. Am. A* **16**(5) (1999) 1175-1183. https://doi.org/10.1364/JOSAA.16.001175

[10] N. C. Lozano-Rincón, J. C. Valencia-Estrada, "Paraboloid-aspheric lenses free of spherical aberration," *J. of M. Optics* **64**(12) (2017) 1146-1157. https://doi.org/10.1080/09500340.2016.1266708

[11] R. G. González-Acuña, H. A. Chaparro-Romo, "General formula for bi-aspheric singlet lens design free of spherical aberration", *Appl. Opt.* **57**(31) (2018) 9341-9345. https://doi.org/10.1364/AO.57.009341

[12] J. Garcia-Marquez, J. C. Valencia-Estrada, H. Perez, S. Topsu, "Catadioptric lenses in visible light communications", *Journal of Physics*: Conf. Series **605** (2015) 012029. https://doi.org/10.1088/1742-6596/605/1/012029

[13] J. C. Valencia-Estrada, R. B. Flores-Hernández, D. Malacara-Hernández, "Singlet lenses free of all orders of spherical aberration", *Proc. R. Soc. A* **471** (2015) 20140608. https://doi.org/10.1098/rspa.2014.0608

[14] K. Schwarzchild, "Untersuchungen zur geometrischen Optik: Einleitung in die Fehlertheorie optischer Instrumente auf Grund des Eikonalbegriffs. I, Druck der Dieterich'schen", Univ. Buchdruckerei, Göttingen (1905).

## MATHEMATICA report
**Difraction limited bi-parabolic lenses**

Clear["Global`*"]

(**Input variables**)
Print["Refractive index        = ",n=1.76]
Print["Object distance         = ",ta=-800]
Print["Image distance          = ",tb=12]
Print["Center thickness        = ",t=0.6]
Print["Lens' diameter          = ",d=5]

(**Output variables**)
m=n-1;
p=n+1;
W=2 fa+m ta;
V=t-n ta;
U=m t ta+2 fa V;

Print["Anterior focal distance   = ",fa=fa/.NSolve[ p U^4+2 n W tb U^3-2n W^2 tb^2 U^2+ n tb^3 (16 fa^3 n^2 ta ((n^2+n-2) ta-fa (n^2-n-2))+8 fa^2 m^2 n^2 ta^3-p t W^4)==0,fa][[4]]]
Print["Back focal distance       = ",fb=(-m U tb)/(2 (U+n W tb))]
Print["Anterior radius           = ",Ra=2 fa]
Print["Back radius               = ",Rb= 2 fb]
Print["Lens' focal distance      = ",F=1/((n-1) (1/(2 fa)-1/(2 fb)+((n-1) t)/(4 fa fb n)))]
Print["F/#                       = ",F/d]
Print["Anterior sagita           = ",d^2/(16fa)]
Print["Thickness complement      = ",t-d^2/(16fa)]

(**Report**)
| | |
|---|---|
| Refractive index | = 1.76 |
| Object distance | = -800 |
| Image distance | = 12 |
| Center thickness | = 0.6 |
| Lens' diameter | = 5 |
| Anterior focal distance | = 5.0973 |
| Back focal distance | = -46.594 |
| Anterior radius | = 10.1946 |
| Back radius | = -93.188 |
| Lens' focal distance | = 12.1216 |
| F/# | = 2.42432 |
| Anterior sagita | = 0.306535 |
| Thickness complement | = 0.293465 |

## OSLO report
**Difraction limited bi-parabolic lenses**

*LENS DATA
| SRF | RADIUS | THICKNESS | APERTURE RADIUS | GLASS | SPECIAL | NOTE |
|---|---|---|---|---|---|---|
| OBJ | 1.0000e+20 | 800.000000 | 100.000000 | AIR | | |
| 1 | 10.194600 | 0.306535 | 2.500000 | GLASS1 | M | * |
| AST | 1.0000e+20 | 0.293465 | 2.500000 AP | GLASS1 | P | |
| 3 | -93.188000 | 12.000000 | 2.500000 P | AIR | | * |
| IMS | -- | -- | 1.538445 S | | | |

*CONIC AND POLYNOMIAL ASPHERIC DATA
| SRF | CC | AD | AE | AF | AG |
|---|---|---|---|---|---|
| 1 | -1.000000 | -- | -- | -- | -- |
| 3 | -1.000000 | -- | -- | -- | -- |

*REFRACTIVE INDICES
| SRF | GLASS | RN1 | TCE |
|---|---|---|---|
| 0 | AIR | 1.000000 | -- |
| 1 | GLASS1 | 1.760000 | 236.000000 |
| 2 | GLASS1 | 1.760000 | 236.000000 |
| 3 | AIR | 1.000000 | 236.000000 |
| 4 | IMAGE SURFACE | | |

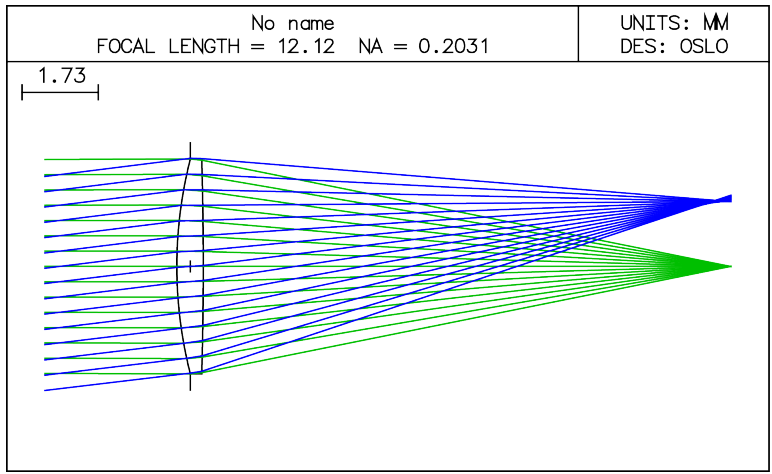
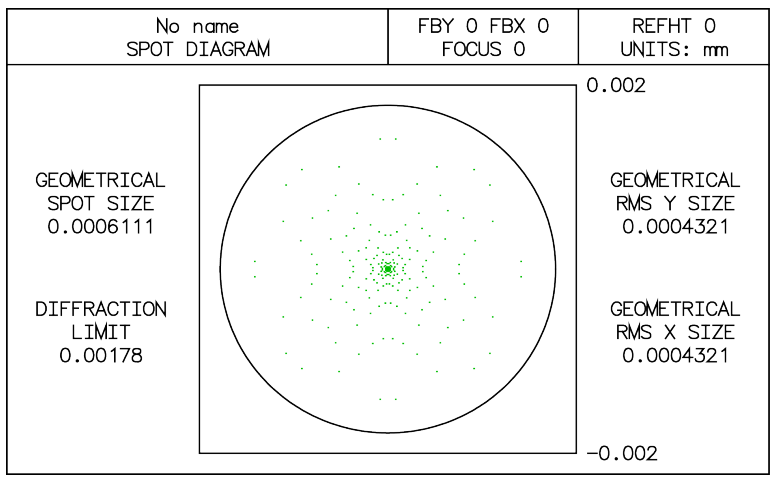